\documentclass[11pt,a4,epsfig]{article}
\usepackage[T1]{fontenc}

\makeatletter

\providecommand{\LyX}{L\kern-.1667em\lower.25em\hbox{Y}\kern-.125emX\@}

\usepackage[T1]{fontenc}

\makeatletter

\usepackage[T1]{fontenc}

\makeatletter

\usepackage[T1]{fontenc}

\makeatletter

\usepackage[T1]{fontenc}

\makeatletter

\usepackage[T1]{fontenc}

\makeatletter

\usepackage[T1]{fontenc}
\usepackage[latin1]{inputenc}

\makeatletter

\usepackage[T1]{fontenc}
\usepackage[latin1]{inputenc}

\makeatletter

\RequirePackage{graphicx}               

\begin{titlepage}
\end{titlepage}
\normalsize
\def\mb#1{\setbox0=\hbox{$#1$}\kern-.025em\copy0\kern-\wd0
\kern-0.05em\copy0\kern-\wd0\kern-.025em\raise.0233em\box0}
\makeatother
\makeatother
\makeatother
\makeatother
\makeatother
\makeatother
\makeatother

\makeatother

\begin{document}

\title{Statistical mechanics of the shallow water system}

\author{P.H. CHAVANIS\protect\protect\( ^{1}\protect \protect \) and J. SOMMERIA\protect\protect\( ^{2}\protect \protect \)\\
\\
 {\small \parbox{11.5cm}{\protect\protect\( ^{1}\protect \protect \)\parbox[t]{11cm}{Laboratoire
de Physique Quantique (CNRS UMR 5626), Universit\'{e} Paul Sabatier, 118 route
de Narbonne 31062 Toulouse, France } \protect\protect\( ^{2}\protect \protect \)\parbox[t]{11cm}{LEGI/CORIOLIS
(CNRS UMR 5519) 21 avenue des Martyrs, 38000 Grenoble, France }}}\small }

\date{\today{}}

\maketitle
We extend the formalism of the statistical theory developed for the 2D Euler
equation to the case of shallow water system. Relaxation equations towards the
maximum entropy state are proposed, which provide a parametrization of sub-grid
scale eddies in 2D compressible turbulence.

\section{Introduction}

\label{sec_introduction}

Two-dimensional flows with high Reynolds numbers have the striking property
of organizing spontaneously into large scale coherent vortices \cite{williams}.
These vortices are common features of geophysical and astrophysical flows with
the well-known example of Jupiter's Great Red Spot, a huge vortex persisting
for more than three centuries in a turbulent shear between two zonal jets. These
vortices share some common features with stellar systems like elliptical or
spherical galaxies that form after a phase of ``violent relaxation'' \cite{lb,csr,cNY}.
They can also have applications in the process of planet formation which may
have begun inside persistent gaseous vortices born out of the protoplanetary
nebula\cite{bs95,tanga,bcps99,chav00}. Understanding and predicting the structure
of these organized states is still a challenging problem.

Since the dynamics of these systems is highly nonlinear, a deterministic description
of the flow for late times is impossible and one must recourse to statistical
methods. The statistical mechanics of 2D flows was first considered by \cite{o49}
Onsager (1949), followed by \cite{jm73} Joyce \& Montgomery (1973), in the
point vortex approximation. The more realistic case of continuous vorticity
fields was later on treated by \cite{k82} Kuzmin (1982),\cite{m90} Miller
(1990) and \cite{rs91} Robert \& Sommeria (1991).

The statistical theory of the Euler equation provides a systematic framework
to tackle the problem of self-organization in 2D flows. At a microscopic level,
complex, inviscid, deformation of the vorticity field creates an intricate filamentation;
however, if we introduce a macroscopic level of description (a ``coarse-graining'')
it can be shown that an overwhelming majority of these microscopic configurations
are close to a macroscopic state (the statistical equilibrium or Gibbs state)
obtained by maximizing a ``mixing entropy'' while accounting for all the constraints
of the Euler equation (the conservation of energy and of the detailed distribution
of vorticity). The maximum entropy theory generally predicts a nonlinear relationship
between vorticity and stream function which respects the properties of the inviscid
dynamics.

The predictions of the statistical theory (\( \nu =0 \)) have been tested in
a large number of numerical simulations and fluid laboratory experiments at
high Reynolds numbers (\( \nu \rightarrow 0 \)). Good agreement is obtained
in many cases, as long as stirring is sufficiently intense to lead to equilibrium
before significant influence of viscous effects \cite{hanna1,hanna}.

However the Euler equation itself is of limited scope for applications to atmospheric
or oceanic motion, where the Coriolis force and density stratification have
a strong influence. A first step in this direction is provided by the quasi-geostrophic
model. The application of the statistical theory to this case is straightforward
as the flow is still assumed non-divergent, and the vorticity is just replaced
by a potential vorticity. This has been discussed in several papers \cite{sndr91,mr94,ksv98,bs00}.

We here extend the theory to the more general case of shallow water system,
a ``compressible'' 2D flow, whose properties are recalled in section \ref{sec_statvelocity}.
Several numerical computations, for instance \cite{di89,pwsf94,cp96}, indicate
inertial organisation of vortical motion into coherent vortices, like with the
incompressible Euler equations. In these flows dominated by vortical motion,
the influence of density waves is weak, and the free surface tends to be controled
by the vortical motion through a balance condition generalizing the quasi-geostrophic
balance.

In section \ref{sec_WV} the procedure of \cite{rs91} Robert and Sommeria (1991)
is extended to the shallow water system with discussion of simplified cases
in section \ref{sec_results}. We still assume that the vorticity field creates
intricate filamentation but the divergence and water height (surface density)
fields are still smooth. These assumptions are justified for flows dominated
by vortical motion at moderate Mach numbers, for which the generation of shocks
is not effective. The relaxation toward equilibrium is discussed in section
\ref{sec_relaxation} providing practical methods for determining the statistical
equilibrium as well as sub-grid scale modeling of turbulence in a shallow water
system. Finally the case of particular geometries is discussed in section \ref{sec_geometries}.

\section{The shallow water equations}

\label{sec_statvelocity}

Consider a fluid layer with thickness \( h(x,y,t) \) submitted to a gravity
\( g \) on a rotating planet. We assume that the layer is thin with respect
to the characteristic length scale of the horizontal motion. In that case, the
velocity field \( {\mathbf{u}}(x,y,t) \) can be assumed two-dimensional and
the vorticity \( {\mb \omega }=\omega {\mathbf{e}_{\mathbf{z}}}=\nabla \wedge {\mathbf{u}} \)
is directed along the vertical axis. We shall assume for simplicity a plane
geometry, with rotation vector \( {\mb \Omega } \) directed along the vertical,
but extension to a spherical geometry would be straigthforward (we introduce
the Coriolis effect but no centrifugal force as the latter is incorporate in
the gravity of the planet). The time evolution of these quantities is determined
by the shallow water equations (see, e.g., \cite{pedlosky} ): 
\begin{equation}
\label{h}
{\partial h\over \partial t}+\nabla \cdot (h{\mathbf{u}})=0
\end{equation}
\begin{equation}
\label{u2}
{\partial {\mathbf{u}}\over \partial t}+({\mb \omega }+2{\mb \Omega })\wedge {\mathbf{u}}=-\nabla B
\end{equation}
 Here the usual advective term \( \mathbf{u}.\nabla \mathbf{u}\textrm{ } \)has
been expressed using the well-known identity of vector analysis \( \mathbf{u}.\nabla \mathbf{u}=\nabla ({{\mathbf{u}}^{2}/2})+{\mb \omega \wedge {\mathbf{u}}} \),
and the term \( \mathbf{u}^{2}/2\textrm{ } \)incorporated in the Bernouilli
function 
\begin{equation}
\label{bernou}
B=gh+{{\mathbf{u}}^{2}\over 2},
\end{equation}
 together with the pressure term \( gh \). Note that the shallow water system
can be viewed as a 2D flow of a compressible gas with density \( h \) and equation
of state \( p=gh^{2}/2 \), and our results could be readily generalized to
2D compressible adiabatic flows. We shall often refer to the shallow water system
as the ``compressible case'', by opposition with the ``incompressible''
case, for which (\ref{h}) is replaced by \( \nabla .\mathbf{u}=0 \) .

One can easily check that the potential vorticity (PV) 
\begin{equation}
\label{PV}
q={\frac{\omega +2\Omega }{h}}
\end{equation}
 is conserved for each fluid parcel, i.e: 
\begin{equation}
\label{dqdt}
{dq\over dt}\equiv {\partial q\over \partial t}+{\mathbf{u}}\cdot \nabla q=0
\end{equation}
 Each mass element \( hd^{2}{\mathbf{r}} \) is conserved during the course
of the evolution. Together with (\ref{dqdt}), this implies the conservation
of the Casimirs 
\begin{equation}
\label{casimir}
C_{f}=\int {f}(q)hd^{2}{\mathbf{r}}
\end{equation}
 where \( f \) is any continuous function of the potential vorticity. In particular,
the moments \( \Gamma _{n} \) of the potential vorticity are conserved 
\begin{equation}
\label{Gn}
\Gamma _{n}=\int q^{n}hd^{2}{\mathbf{r}}
\end{equation}
 The moments \( n=0,1,2 \) are respectively the total mass \( M \), the circulation
\( \Gamma  \) and the PV enstrophy \( \Gamma _{2} \). The energy 
\begin{equation}
\label{energy}
E=\int h{{\mathbf{u}}^{2}\over2 }d^{2}{\mathbf{r}}+{1\over 2}\int gh^{2}d^{2}{\mathbf{r}}
\end{equation}
 involving a kinetic and potential part, is also a conserved quantity. Note
finally that for a multiply connected domain, like the annulus or channel discussed
in section \ref{sec_geometries}, the circulation along each boundary is conserved,
in addition to \( \Gamma  \) .

It will be convenient in the sequel to use a Helmholtz decomposition of the
momentum \( h{\mathbf{u}} \) into a purely rotational and a purely divergent
part 
\begin{equation}
\label{hu}
h{\mathbf{u}}=-{\mathbf{e}}_{z}\wedge \nabla \psi +\nabla \phi 
\end{equation}
 where \( \psi  \) and \( \phi  \) are defined as solutions of the Poisson
equations 
\begin{equation}
\label{psi}
\Delta \psi =-\nabla \wedge (h{\mathbf{u}}),\qquad \psi =const.\quad {\textrm{on each boundary}}
\end{equation}
\begin{equation}
\label{phi}
\Delta \phi =-\nabla \cdot (h{\mathbf{u}}),\qquad \partial \phi /\partial \zeta =0\quad {\textrm{on each boundary}}
\end{equation}
 where the conditions at the domain boundary (with normal coordinate \( \zeta \textrm{ } \))
are the consequences of the impermeability condition. We here consider a domain
with a single (outer) boundary, so we can take \( \psi =0 \) at this boundary
(as \( \psi  \) is defined within an arbitrary gauge constant).

For a steady solution, the mass conservation (\ref{h}) reduces to \( \nabla \cdot (h{\mathbf{u}})=0 \),
so that \( \phi =0 \) and 
\begin{equation}
\label{husteady}
h{\mathbf{u}}=-{\textbf {e}}_{z}\wedge \nabla \psi \qquad ({\textrm{steady}}),
\end{equation}
 and equation (\ref{dqdt}) reduces to \( {\mathbf{u}}\cdot \nabla q=0 \),
which implies that \( q\textrm{ } \)is a function \( F\textrm{ } \)of the
stream function \( \psi  \)
\begin{equation}
\label{qpsige}
q=F(\psi )
\end{equation}
 Finally, equation (\ref{u2}) reduces to 
\begin{equation}
\label{bernsteady}
({\mb \omega }+2{\mb \Omega })\wedge {\mathbf{u}}=-\nabla B
\end{equation}
 Taking the dot product with \( {\mathbf{u}} \), we obtain \( {\mathbf{u}}\cdot \nabla B=0 \)
or, equivalently, \( B=B(\psi ) \). This is known as the Bernouilli theorem.
Substituting for (\ref{husteady}) in equation \( (\ref {bernsteady}) \), we
obtain a specific relationship between the potential vorticity \( q\textrm{ } \)and
the Bernouilli function \( B\textrm{ } \)in the form: 
\begin{equation}
\label{dBdpsi}
q=-{dB\over d\psi }
\end{equation}

Small perturbations to a state of rest, with uniform thickness \( H\textrm{ } \),
satisfy linearized equations with two branches of solutions. For small scales,
these are the usual surface waves on one hand, with purely divergent velocity
and propagation speed \( c=(gH)^{1/2} \), and steady vortical divergenceless
modes on the other hand. In nonlinear regimes, these two modes interact. Vortical
motion with scale \( l\textrm{ and } \)typical vorticity \( \omega  \) fluctuates
on time scale \( \omega ^{-1\textrm{ }} \), so it emits waves at wavelength
\( \lambda \sim c/\omega  \), hence \( \lambda /l \) is the inverse of the
Mach number \( c/u \) based on the local velocity \( u\sim \omega l \). Therefore
we expect that for the case of small Mach numbers that we shall consider, velocity
divergence and free surface deformation are much smoother than the vorticity
field (their wavelength is much larger).

\section{The maximum entropy theory}

\label{sec_WV}

\subsection{General principles and notations}

\label{sec_notations}

For slow velocities, the shallow water system reduces to the quasi-geostrophic
(QG) equations, with \( h\simeq cte\textrm{ } \), such that (\ref{h}) reduces
to the incompressibility condition \( \nabla .\mathbf{u}=0\textrm{ } \). Then
the velocity field remains regular for any time, but it generally develops complex
fine scale vorticity filaments so that a deterministic description of the flow
would require a rapidly increasing amount of information as time goes on. The
idea of the statistical theory is to give up \textit{\emph{such a}} \textit{deterministic}
description and refer to a \textit{probabilistic} description. Therefore, the
exact knowledge of the ``fine-grained'', or microscopic potential vorticity
field is replaced by the probability density (area fraction) \( \rho ({\mathbf{r}},\sigma ) \)
of finding the potential vorticity level \( \sigma  \) at position \( {\mathbf{r}} \).

For the more general shallow water system, the inviscid dynamics generally leads
to singularities (shocks), with associated energy dissipation (even in the absence
of viscosity). This is a source of fundamental mathematical difficulty for the
generalization of the equilibrium statistical mechanics initially developed
for the Euler equations or QG system. Nevertheless, for the case of small Mach
numbers that we consider, shocks occur only after a very long time (due to non-linear
steepening of surface waves), and they involve a weak energy dissipation, since
most of the energy remains in the vortical motion. Furthermore fine scale vorticity
fluctuations behave like in the QG system, and only interact with surface waves
and flow divergence at much larger scale, as discussed above. We shall therefore
assume that vorticity fluctuates at small scale, but divergence is smooth as
well as the height \( h \).

We still describe the local vorticity fluctuations by the probability \( \rho ({\mathbf{r}},\sigma ) \)
of finding the potential vorticity level \( \sigma  \) in a small neighborhood
of the position \( {\mathbf{r}} \). The normalization condition yields at each
point 
\begin{equation}
\label{norm}
\int \rho ({\mathbf{r}},\sigma )d\sigma =1
\end{equation}
 The locally averaged field of potential vorticity is expressed in terms of
the probability density in the form 
\begin{equation}
\label{qbar}
\overline{q}=\int \rho ({\mathbf{r}},\sigma )\sigma d\sigma 
\end{equation}
 This locally averaged field is called the macroscopic, or coarse-grained, potential
vorticity. A macroscopic state is fully defined by \( \rho ({\mathbf{r}},\sigma ) \),
the height field \( h(\mathbf{r}) \) and the flow divergence (assumed without
microscopic fluctuations). The velocity field \( \mathbf{u}(\mathbf{r}) \)
can be also considered as smooth, as it integrates the vorticity fluctuations,
and can be deduced by integration from its divergence and vorticity \( \overline{\omega }=\overline{q}h-2\Omega  \)
. The energy (\ref{energy}) depends only on this smooth field, with negligible
influence of local fluctuations, like in the incompressible case \cite{rs91}.

The conservation of the Casimirs (\ref{casimir}) is equivalent to the conservation
of the global distribution of potential vorticity (i.e the total area of each
level of potential vorticity ponderated by \( h \)): 
\begin{equation}
\label{gamma}
\gamma (\sigma )=\int \rho ({\mathbf{r}},\sigma )h({\mathbf{r}})d^{2}{\mathbf{r}}
\end{equation}
 The microscopic moments of potential vorticity can be written: 
\begin{equation}
\label{Gamma}
\Gamma _{n}=\int \gamma (\sigma )\sigma ^{n}d\sigma =\int \overline{q^{n}}h({\mathbf{r}})d^{2}{\mathbf{r}}
\end{equation}
 where 
\begin{equation}
\label{qn}
\overline{q^{n}}=\int \rho ({\mathbf{r}},\sigma )\sigma ^{n}d\sigma 
\end{equation}
 and are conserved during an inviscid evolution. Note that for \( n\geq 2 \),
the macroscopic moments of potential vorticity \( \Gamma ^{c.g.}_{n}=\int \overline{q}^{n}h({\mathbf{r}})d^{2}{\mathbf{r}} \)
are \textit{not} conserved, as there are partly transfered into fine-grained
fluctuations.

The mixing entropy 
\begin{equation}
\label{entropy}
S=-\int \rho ({\mathbf{r}},\sigma )\ln \rho ({\mathbf{r}},\sigma )h({\mathbf{r}})d^{2}{\mathbf{r}}d\sigma 
\end{equation}
 measures the number of microscopic configurations associated with the same
macroscopic field of potential vorticity. The dependence in \( \rho  \) is
the same as for the incompressible case \cite{rs91}, and the factor \( h({\mathbf{r}}) \)
is introduced to insure that entropy is conserved by mere macroscopic displacement
of fluid parcels. Indeed the mass element \( h({\mathbf{r}})d^{2}{\mathbf{r}} \)
is conserved by fluid particle displacement instead of the surface element \( d^{2}{\mathbf{r}} \)
in the incompressible case. At equilibrium, the system is expected to be in
the most probable (i.e. most mixed) state consistent with the constraints of
the Euler equation. The entropy (\ref{entropy}) has been justified by rigorous
mathematical arguments (in the incompressible case) but other forms have been
recently proposed (see discussion in \cite{hanna}). Therefore, we shall consider
a general form of entropy 
\begin{equation}
\label{entrgen}
S=-\int s(\rho ({\mathbf{r}},\sigma ))h({\mathbf{r}})d^{2}{\mathbf{r}}d\sigma 
\end{equation}
 and find that equation (\ref{entropy}) is the only expression leading to a
well defined entropy extremum.

\subsection{First order variations}

\label{sec_variations}

According to the previous discussion, the most probable macroscopic state is
obtained by maximizing the entropy (\ref{entrgen}) with fixed energy (\ref{energy}),
global vorticity distribution (\ref{gamma}) and local normalization (\ref{norm}).
This problem is treated by introducing Lagrange multipliers, so that the first
variations satisfy 
\begin{equation}
\label{varia}
\delta S-\beta \delta E-\int \alpha (\sigma )\delta \gamma (\sigma )d\sigma -\int \zeta ({\mathbf{r}})\delta \biggl (\int \rho ({\mathbf{r}},\sigma )d\sigma \biggr )hd^{2}{\mathbf{r}}=0.
\end{equation}
 The Lagrange multipliers are respectively the ``inverse temperature'' \( \beta , \)
the ``chemical potential'' \( \alpha (\sigma ) \) of species \( \sigma , \)
and \( \zeta ({\mathbf{r}}) \).

We shall take \( h \), \( \rho h \) and \( \nabla \cdot \mathbf{u} \) as
independent variables characterizing the macroscopic state. Then, it is easy
to establish, by differentiating respectively (\ref{entropy}), (\ref{gamma})
and (\ref{norm}), that: 
\begin{equation}
\label{dS}
\delta S=\int \lbrack \rho s'(\rho )-s(\rho )\rbrack \delta hd^{2}{\mathbf{r}}d\sigma -\int s'(\rho )\delta (\rho h)d^{2}{\mathbf{r}}d\sigma 
\end{equation}
\begin{equation}
\label{dgamma}
\delta \gamma (\sigma )=\int \delta (\rho h)d^{2}{\mathbf{r}}
\end{equation}
\begin{equation}
\label{d1}
h\delta \biggl (\int \rho ({\mathbf{r}},\sigma )d\sigma \biggr )=\int \delta (\rho h)d\sigma -\int {\rho }\delta hd\sigma 
\end{equation}
 The variations of the energy are conveniently expressed in terms of the Bernouilli
function \( B \), 
\begin{equation}
\label{dE1}
\delta E=\int B\delta hd^{2}{\mathbf{r}}+\int h\mathbf{u}\cdot \delta \mathbf{u}d^{2}{\mathbf{r}}
\end{equation}
 Then, using the Helmholtz decomposition (\ref{hu}) for the momentum \( h\mathbf{u} \)
, the second integral can be rewritten 
\begin{equation}
\label{dE2}
\int h\mathbf{u}\cdot \delta \mathbf{u}d^{2}{\mathbf{r}}=-\int (\nabla \psi \wedge \delta \mathbf{u})\cdot {\mathbf{z}}d^{2}{\mathbf{r}}+\int \nabla \phi \cdot \delta \mathbf{u}d^{2}{\mathbf{r}}
\end{equation}
 Integrating by parts with the identities \( \nabla \wedge (\psi \delta \mathbf{u})=\psi \nabla \wedge (\delta \mathbf{u})+\nabla \psi \wedge \delta \mathbf{u} \)
and \( \nabla \cdot (\phi \delta \mathbf{u})=\phi \nabla \cdot (\delta \mathbf{u})+\nabla \phi \cdot \delta \mathbf{u} \),
and using the boundary conditions for \( \psi  \) and \( \phi  \) , we obtain
\begin{equation}
\label{dE3}
\int h\mathbf{u}\cdot \delta \mathbf{u}d^{2}{\mathbf{r}}=\int \psi \delta \overline{\omega }d^{2}{\mathbf{r}}-\int \phi \nabla \cdot (\delta \mathbf{u})d^{2}{\mathbf{r}}
\end{equation}
 Using (\ref{norm}) (\ref{PV}) and (\ref{qbar}), we have finally 
\begin{equation}
\label{dE4}
\delta E=\int {B}\rho \delta hd^{2}{\mathbf{r}}d\sigma +\int \psi \sigma \delta (\rho h)d^{2}{\mathbf{r}}d\sigma -\int \phi \delta (\nabla \cdot \mathbf{u})d^{2}{\mathbf{r}}
\end{equation}

The variation (\ref{varia}) vanishes for any small changes of the variables
only if the coefficient of each independent variable vanishes: 
\begin{equation}
\label{var1}
\delta (\rho h):\qquad s'(\rho )=-\beta \sigma \psi -\alpha (\sigma )-{\zeta ({\mathbf{r}})}
\end{equation}
\begin{equation}
\label{var2}
\delta h:\qquad s'(\rho )-{s(\rho )\over \rho }=\beta B-{\zeta ({\mathbf{r}})}
\end{equation}
\begin{equation}
\label{var3}
\delta (\nabla \cdot \mathbf{u}):\qquad \phi =0\qquad \qquad \qquad \quad 
\end{equation}
 Note that the right hand side of equation (\ref{var2}) is independent of \( \sigma  \).
This implies that the term on the left hand side must be a constant (that we
can take equal to \( 1 \) without loss of generality):\( s'(\rho )-{{s(\rho )/\rho }}=1 \).
This equation is easily integrated in \( s(\rho )=A\rho +\rho \ln \rho  \)
where \( A \) is an integration constant. When substituted in equation (\ref{entrgen}),
using (\ref{Gamma}), this yields 
\begin{equation}
\label{func}
S=-\int \rho ({\mathbf{r}},\sigma )\ln \rho ({\mathbf{r}},\sigma )h({\mathbf{r}})d^{2}{\mathbf{r}}d\sigma -AM
\end{equation}
 which is just the entropy (\ref{entropy}) up to an additive constant term
\( AM \) (which we can take equal to zero without loss of generality). Therefore,
the entropy (\ref{entropy}) is the only functional of the form (\ref{entrgen})
for which the maximization problem has a solution. This result is astounding
because it is obtained without any explicit reference to thermodynamical arguments.

\subsection{The Gibbs states}

\label{sec_gibbs}

Substituting explicitely \( s(\rho )=\rho \ln \rho  \) in equation (\ref{var1}),
the optimal probability density can be expressed as 
\begin{equation}
\label{rhopt}
\rho ({\mathbf{r}},\sigma )={1\over Z(\psi )}g(\sigma )e^{-\beta \sigma \psi }
\end{equation}
 where \( Z(\psi )\equiv e^{{\zeta ({\mathbf{r}})}+1} \) and \( g(\sigma )\equiv e^{-\alpha (\sigma )} \).
The normalization condition (\ref{norm}) leads to a value of the partition
function \( Z \) of the form 
\begin{equation}
\label{Z}
Z=\int g(\sigma )e^{-\beta \sigma \psi }d\sigma 
\end{equation}
 and the locally averaged potential vorticity (\ref{qbar}) is expressed as
a function of \( \psi  \) according to: 
\begin{equation}
\label{qpsiexpli}
\overline{q}={\int g(\sigma )\sigma e^{-\beta \sigma \psi }d\sigma \over \int g(\sigma )e^{-\beta \sigma \psi }d\sigma }=F(\psi )
\end{equation}
 This can be rewritten 
\begin{equation}
\label{qpsi}
\overline{q}=-{1\over \beta }{d\ln Z\over d\psi }
\end{equation}
 This is the same functional relation as in the case of 2D incompressible Euler
flows \cite{rs91}.

Differentiating equation (\ref{qpsiexpli}) with respect to \( \psi  \), we
check that the variance of the potential vorticity can be written 
\begin{equation}
\label{q2fprime}
q_{2}\equiv \overline{q^{2}}-\overline{q}^{2}=-{1\over \beta }F'(\psi )
\end{equation}
 or, alternatively (see equation (\ref{qpsi})): 
\begin{equation}
\label{q2mq2}
q_{2}={1\over \beta ^{2}}{d^{2}\ln Z\over d\psi ^{2}}
\end{equation}
 Therefore the slope of the function \( \overline{q}=F(\psi ) \) is directly
related to the variance of the vorticity distribution. Relation (\ref{q2fprime})
has the same form and origin as the ``fluctuation-dissipation'' theorem in
statistical field theory, where \( d\overline{q}/d\psi  \) is interpreted as
a susceptibility. Since \( q_{2}>0 \), we find that the function \( \overline{q}=F(\psi ) \)
is monotonic; it is decreasing for \( \beta >0 \) and increasing for \( \beta <0 \)
(it is constant for \( \beta =0 \)). Another proof of this result is given
in \cite{rs91}.

Substituting explicitely \( s'(\rho )-s(\rho )/\rho =1 \) in equation (\ref{var2}),
we have 
\begin{equation}
\label{Bpsi}
B={1\over \beta }\ln Z
\end{equation}
 This relation shows that the Bernouilli function plays the role of a free energy
in the statistical theory. We note that both \( B \) and \( \overline{q} \)
are functions of \( \psi  \), while \( \phi =0 \) , from (\ref{var3}), as
it should for steady flows. Furthermore, taking the derivative of (\ref{Bpsi})
with respect to \( \psi  \) and using (\ref{qpsi}), we check that the relation
\( \overline{q}=-dB/d\psi  \) required for a steady solution of the shallow
water equation is satisfied. Therefore the flow selected by the purely statistical
entropy maximization procedure does not evolve anymore by the flow evolution,
so the statistical theory is indeed consistent with the dynamics.

\section{Properties of the Gibbs states in some particular cases:}

\label{sec_results}

\subsection{Particular \protect\protect\protect\protect\( \overline{q}-\psi \protect \protect \protect \protect \)
relationships:}

\label{sec_particular}

The Gibbs states are characterized by the relation (\ref{qpsiexpli}) between
\( \overline{q} \) and \( \psi  \), which is always monotonic, as shown in
the previous section. It is determined by the conservation laws, but only indirectly
through the set of Lagrange multipliers \( \beta  \) and \( \alpha (\sigma ) \).
In practice we need to discretize the PV levels, and keeping only two levels,
\( q=\sigma _{0} \) and \( q=\sigma _{1} \) is often representative of more
general cases. Then the probability distribution \( \rho  \) just depends on
a single probability \( p_{1} \) of finding the level \( \sigma _{1} \) (for
instance), with the probability \( 1-p_{1} \) of finding the complementary
level \( \sigma _{0} \) , i.e. \( g(\sigma ) \) is the sum of two Dirac function
terms, \( g(\sigma )=g_{1}[\lambda \delta (\sigma -\sigma _{0})+\delta (\sigma -\sigma _{1})] \).
This probability \( p_{1} \) is directly related to the PV average by \( \overline{q}=p_{1}\sigma _{1}+(1-p_{1})\sigma _{0} \),
or reversely \( p_{1}=(\overline{q}-\sigma _{0})/(\sigma _{1}-\sigma _{0}) \).
The expression (\ref{qpsiexpli}) for \( \overline{q} \) reduces to 
\begin{equation}
\label{FD}
\overline{q}=\sigma _{0}+{(\sigma _{1}-\sigma _{0})\over 1+\lambda e^{\beta (\sigma _{1}-\sigma _{0})\psi }}
\end{equation}
 This relation corresponds to the Fermi-Dirac statistics. The two unknown parameters
\( \lambda  \) and \( \beta  \) are indirectly determined by the conserved
quantities. The associated Bernouilli function (\ref{Bpsi}) becomes 
\begin{equation}
\label{BFD}
B={1\over \beta }\ln g_{1}-\sigma _{0}\psi +{1\over \beta }\ln \biggl \lbrace \lambda +e^{\beta (\sigma _{0}-\sigma _{1})\psi }\biggr \rbrace 
\end{equation}

The problem is also greatly simplified in the alternative case for which \( g(\sigma ) \)
is a Gaussian: 
\begin{equation}
\label{gg}
g(\sigma )=g_{0}e^{-(\sigma -\sigma _{*})^{2}\over 2\sigma _{2}}.
\end{equation}
 Then the local probability distribution (\ref{rhopt}) is also a Gaussian,
and the corresponding Bernouilli function (\ref{Bpsi}) is 
\begin{equation}
\label{Blin}
B={1\over \beta }\ln [g_{0}(2\pi \sigma _{2})^{1/2}]+{1\over 2}\sigma _{2}\beta \psi ^{2}-\sigma _{*}\psi ,
\end{equation}
 corresponding to a linear relationship: 
\begin{equation}
\label{qpsilin}
\overline{q}=-\beta \sigma _{2}\psi +\sigma _{*}
\end{equation}
 According to (\ref{q2fprime}) the variance of the potential vorticity is then
uniform, with value \( q_{2}=\sigma _{2} \)(more generally, all the even momenta
of the Gaussian are related to \( \sigma _{2} \) by \( \overline{(q-\overline{q})^{2n}}=(2n-1)!!\sigma _{2}^{n} \)
and the odd momenta cancel).

This Gaussian local probability distribution is obtained by maximizing the entropy
(\ref{entropy}), reducing the constraints of the global distribution \( \gamma (\sigma ) \)
to its first moments \( \Gamma _{0}\equiv M \), \( \Gamma _{1} \) and \( \Gamma _{2} \).
This will be the true Gibbs state for a particular initial distribution \( \gamma (\sigma ) \)
with higher order moments equal to the global moments of this simplified Gibbs
state. A linear relationship between \( \overline{q} \) and \( \psi  \) can
also be obtained for any distribution \( \gamma (\sigma ) \) in the limit of
strong mixing where \( \beta \sigma \psi \ll 1 \), so that (\ref{qpsiexpli})
can be linearized, as discussed in \cite{cs96} Chavanis \& Sommeria (1996).

\subsection{Unidirectional solutions}

\label{sec_jets}

We consider here unidirectional solutions such that \( \mathbf{u}=u(y){\mathbf{e}}_{x} \).
The equilibrium relation \( \overline{q}=-dB/d\psi  \) then yields, multiplying
each member by \( hu=d\psi /dy \), 
\begin{equation}
\label{premiere}
g{dh\over dy}+{2\Omega \over h}{d\psi \over dy}=0.
\end{equation}
 This condition of geostrophic balance can be readily integrated in 
\begin{equation}
\label{premiereint}
h=H\sqrt{1-{4\Omega \over gH^{2}}\psi }
\end{equation}
 A second relation is provided by the expression of the Bernouilli function
yielding 
\begin{equation}
\label{Bernouilliuni}
{1\over 2h^{2}}\biggl ({d\psi \over dy}\biggr )^{2}=B(\psi )-gh
\end{equation}
 Combined with the expression (\ref{premiereint}) of \( h \) and the Gibbs
state expression (\ref{Bpsi}) of \( B(\psi ) \), this yields a first order
ordinary differential equation for \( \psi . \) This equation depends on the
constant \( H \) and Lagrange parameters, for instance \( g_{1}, \) \( \beta  \)
and \( \lambda  \) in the case with two PV levels. The constraints on total
mass \( M \), global mass of PV level \( \sigma _{1} \) and total energy \( E \),
indirectly determine these parameters. Note that in reality this solution must
be viewed in a x-wise translating frame of reference, as discussed in section
\ref{sec_geometries}, due to the additional conservation law for momentum.

\subsection{Axisymmetric solutions}

\label{sec_axi}

For axisymmetric solutions \( {\mathbf{u}}=u_{\theta }(r){\mathbf{e}}_{\theta } \)
where \( (r,\theta ) \) are polar coordinates, and \( hu_{\theta }=-d\psi /dr \).
Then Equation (\ref{premiere}) is replaced by the cyclostrophic balance 
\begin{equation}
\label{first}
gh{dh\over dr}={1\over hr}{\biggl ({d\psi \over dr}\biggr )^{2}}-2\Omega {d\psi \over dr}
\end{equation}
 When \( hu_{\theta }=-d\psi /dr\geq 0 \) (cyclone), \( h \) is an increasing
function of \( r \), so the vortex core is a trough. In the opposite case \( u_{\theta }\leq 0 \)
(anticyclone) , the vortex core is a bump in geostrophic regimes. However for
large velocities (Rossby number larger than unity), the term \( {u_{\theta }^{2}\over r} \)
dominates, leading always to a trough.

The expression of the Bernouilli constant gives again the form (\ref{Bernouilliuni}),
just replacing \( y \) by the radius \( r \) . Combining this relation with
(\ref{first}), one gets a couple of first order ordinary differential equations
in the variables \( \psi  \) and \( h. \) As in the unidirectional case, the
solution depends on two constants of integration and the Lagrange parameters,
which are \( g_{1}, \) \( \beta  \) and \( \lambda  \) in the case with two
PV levels. This solution must be viewed in general in a rotating frame of reference,
due to the additional conservation of angular momentum , as discussed in section
\ref{sec_geometries}.

\section{Relaxation equations}

\label{sec_relaxation}

\subsection{The Maximum Entropy Production Principle}

\label{sec_MEPP}

Relaxation methods are convenient to compute the statistical equilibrium resulting
from any initial condition. The aim is to increase the entropy in successive
steps while keeping constant all the conserved quantities. \cite{tw96} Turkington
\& Whitaker (1996) have implemented a relaxation method to calculate the Gibbs
states obtained with the Euler equations. \cite{rs92} Robert and Sommeria (1992)
had previously proposed relaxation equations in the form of a parameterization
of sub-grid scale eddies which drives the system toward statistical equilibrium
by a continuous time evolution. Such relaxation equations can be used both as
a realistic coarse resolution model of the turbulent evolution, and as a method
of determination of the statistical equilibrium resulting from this evolution
(see \cite{csr} for a short review). We here generalize this approach to the
shallow water system.

We first decompose the vorticity \( \omega  \) and velocity \( {\mathbf{u}} \)
into a mean and fluctuating part, namely \( \omega =\overline{\omega }+\tilde{\omega } \),
\( {\mathbf{u}}=\overline{\mathbf{u}}+\tilde{\mathbf{u}} \), keeping \( h \)
smooth. Taking the local average of the shallow water equations (\ref{h})(\ref{u2}),
we get 
\begin{equation}
\label{hbar}
{\partial h\over \partial t}+\nabla \cdot (h\overline{\mathbf{u}})=0
\end{equation}
\begin{equation}
\label{u2bar}
{\partial \overline{\mathbf{u}}\over \partial t}+(\overline{\mb \omega }+2{\mb \Omega })\wedge \overline{\mathbf{u}}=-\nabla B-{\mathbf{e}}_{z}\wedge {\mathbf{J}}_{\omega }
\end{equation}
\begin{equation}
\label{bernoubar}
B=gh+{\overline{\mathbf{u}}^{2}\over 2}
\end{equation}
 where the current \( {\mathbf{J}}_{\omega }=\overline{\tilde{\omega }\tilde{\mathbf{u}}} \)
represents the correlations of the fine-grained fluctuations. Although we have
neglected the fluctuation energy \( \tilde{\mathbf{u}}^{2} \) in front of \( \overline{\mathbf{u}}^{2} \)
(as well as fluctuations of \( h \)), we keep the correlations \( {\mathbf{J}}_{\omega }=\overline{\tilde{\omega }\tilde{\mathbf{u}}} \),
which represent the PV transport by sub-grid-scale eddies. This assumption is
justified since, denoting \( \epsilon  \) the typical scale of vorticity fluctuations,
we have \( \tilde{\mathbf{u}}^{2}\sim \epsilon ^{2}\overline{\omega }^{2} \)
and \( \overline{\tilde{\omega }\tilde{\mathbf{u}}}\sim \epsilon \overline{\omega }^{2}\gg \tilde{\mathbf{u}}^{2} \)
(while \( \tilde{\omega }\sim \overline{\omega } \)).

We deduce an equation for the evolution of the potential vorticity (\ref{PV}),
taking the curl of (\ref{u2bar}) and using (\ref{hbar}), 
\begin{equation}
\label{hq}
{\partial \over \partial t}(h\overline{q})+\nabla .(h\overline{q}\; \overline{\mathbf{u}})=-\nabla .{\mathbf{J}}_{\omega }
\end{equation}
 This equation can be viewed as a local conservation law for the circulation
\( \Gamma =\int \overline{q}hd^{2}{\mathbf{r}} \). We shall determine the unknown
current \( {\mathbf{J}}_{\omega } \) by the thermodynamic principle of Maximum
Entropy Production (MEP) \cite{rs92}. For that purpose, we need to consider
not only the locally average PV field \( \overline{q} \), but the whole probability
distribution \( \rho ({\mathbf{r}},{\sigma },t) \) now evolving with time \( t \).
The conservation of the global vorticity distribution \( \gamma (\sigma )=\int \rho hd^{2}{\mathbf{r}} \)
can be written in the local form 
\begin{equation}
\label{rhot}
{\partial \over \partial t}(h\rho )+\nabla .(h\rho \overline{\mathbf{u}})=-\nabla .{\mathbf{J}}
\end{equation}
 where \( {\mathbf{J}}({\mathbf{r}},{\sigma },t) \) is the (unknown) current
associated with the level \( \sigma  \) of potential vorticity. Integrating
equation (\ref{rhot}) over all the PV levels \( \sigma  \), using (\ref{norm}),
and comparing with (\ref{hbar}), we find the constraint 
\begin{equation}
\label{Jnorm}
\int {\mathbf{J}}({\mathbf{r}},\sigma ,t)d\sigma =0
\end{equation}
 Multiplying (\ref{rhot}) by \( \sigma  \), integrating over all the PV levels,
using (\ref{qbar}) and comparing with (\ref{hq}), we get 
\begin{equation}
\label{Jsigma}
\int {\mathbf{J}}({\mathbf{r}},\sigma ,t)\sigma d\sigma ={\mathbf{J}}_{\omega }
\end{equation}

We can express the time variation of the energy \( \dot{E}\equiv dE/dt \) in
terms of \( {\mathbf{J}} \), using (\ref{energy}) and (\ref{u2bar}), leading
to the energy conservation constraint 
\begin{equation}
\label{Edot}
\dot{E}=\int {\mathbf{J}}\sigma h\overline{\mathbf{u}}_{\perp }d^{2}{\mathbf{r}}d\sigma =0\; ,
\end{equation}
 where \( \overline{\mathbf{u}}_{\perp }\equiv {\mathbf{e}}_{z}\wedge \overline{\mathbf{u}} \)
. Using (\ref{entropy}) and (\ref{rhot}), we similarly express the rate of
entropy production as 
\begin{equation}
\label{Sdot}
\dot{S}=-\int {\mathbf{J}}\nabla (\ln \rho )d^{2}{\mathbf{r}}d\sigma \; .
\end{equation}

According to the Maximum Entropy Production Principle, we determine the current
\( {\mathbf{J}} \) which maximizes the rate of entropy production \( \dot{S} \)
respecting the constraints \( \dot{E}=0 \), (\ref{Jnorm}) and \( \int {J^{2}\over 2\rho }d\sigma \leq C({\mathbf{r}},\sigma ) \).
The last constraint expresses a bound (unknown) on the value of the diffusion
current. Convexity arguments justify that this bound is always reached so that
the inequality can be replaced by an equality. The corresponding condition on
first variations can be written at each time \( t \): 
\begin{equation}
\label{vp}
\delta \dot{S}-\beta (t)\delta \dot{E}-\int \zeta ({\mathbf{r}},t)\delta \biggl (\int {\mathbf{J}}d\sigma \biggr )d^{2}{\mathbf{r}}-\int D^{-1}({\mathbf{r}},t)\delta \biggl (\int {J^{2}\over 2\rho }\biggr )d\sigma d^{2}{\mathbf{r}}=0
\end{equation}
 and leads to a current of the form 
\begin{equation}
\label{Jopt}
{\mathbf{J}}=-D({\mathbf{r}},t)(\nabla \rho +\beta (t)\sigma \rho h\overline{\mathbf{u}}_{\perp }-\zeta ({\mathbf{r}},t)\rho )
\end{equation}
 The Lagrange multiplier \( \zeta ({\mathbf{r}},t) \) is determined by the
constraint (\ref{Jnorm}), which leads to 
\begin{equation}
\label{Jopt2}
{\mathbf{J}}=-D({\mathbf{r}},t)\biggl \lbrack \nabla \rho +\beta (t)\rho (\sigma -\overline{q})h\overline{\mathbf{u}}_{\perp }\biggr \rbrack 
\end{equation}
 This optimal current is similar to the one obtained in ordinary incompressible
2D turbulence except that the term \( h\overline{\mathbf{u}}_{\perp } \) now
replaces \( \nabla \psi . \) The impermeability condition imposes that the
normal component of the velocity and of the current vanishes at the wall. We
therefore have the boundary conditions 
\begin{equation}
\label{uboundary}
{\textbf {n}}\overline{.\textbf {u}}=0\qquad ({\textrm{on }\, \, \textrm{each }\, \, \textrm{boundary}})
\end{equation}
\begin{equation}
\label{Jboundary}
{\textbf {n}}.\nabla \rho =-\beta (t)\rho (\sigma -\overline{q})h{\textbf {n}}\overline{\mathbf{u}}_{\perp }\qquad ({\textrm{on }\, \, \textrm{each }\, \, \textrm{boundary}})
\end{equation}
 where \( {\textbf {n}} \) is a unit vector normal to the wall.

The diffusion coefficient \( D \) is not determined by the MEPP as it depends
on the unknown bound \( C \) on the current. It can be determined by more systematic
procedures inspired from kinetic theories of plasma physics like in \cite{c98,c00}
for the incompressible case. For the purpose of reaching the Gibbs state, it
can simply be chosen arbitrarily. However, we shall show below that \( D \)
must be positive so as to insure entropy increase.

The conservation of energy (\ref{Edot}) at any time determines the evolution
of the Lagrange multiplier \( \beta (t) \) according to 
\begin{equation}
\label{betat}
\beta (t)=-{\int D\nabla \overline{q}h\overline{\mathbf{u}}_{\perp }d^{2}{\mathbf{r}}\over \int D(\overline{q^{2}}-\overline{q}^{2})(h\overline{\mathbf{u}}_{\perp })^{2}d^{2}{\mathbf{r}}}
\end{equation}

We can now provide an explicit form for the vorticity current \( {\mathbf{J}}_{\omega }\textrm{ } \)to
introduce back in the shallow water equation (\ref{u2bar}). Indeed, using (\ref{Jopt2})
and (\ref{qbar}), we find 
\begin{equation}
\label{Jomega}
{\mathbf{J}}_{\omega }=-D\biggl \lbrack \nabla \overline{q}+\beta (t)(\overline{q^{2}}-\overline{q}^{2})h\overline{\mathbf{u}}_{\perp }\biggr \rbrack 
\end{equation}
 Substituting (\ref{Jomega}) in equation (\ref{u2bar}), we obtain 
\begin{equation}
\label{ubar3}
{\partial \overline{\mathbf{u}}\over \partial t}+(\overline{\mb \omega }+2{\mb \Omega })\wedge \overline{\mathbf{u}}=-\nabla B+D\biggl \lbrack {\mathbf{e}}_{z}\wedge \nabla \overline{q}-\beta (t)(\overline{q^{2}}-\overline{q}^{2})h\overline{\mathbf{u}}\biggr \rbrack 
\end{equation}
 Since \( \beta (t)\leq 0 \) in relevant situations, the last term in (\ref{ubar3})
represents a \textit{forcing} proportional to \( \overline{\mathbf{u}} \) which
compensates the diffusion caused by the term \( {\hat{\mathbf{e}}}_{z}\wedge \nabla \overline{q}\sim \Delta \overline{\mathbf{u}} \).
This additional term depends on the local PV variance \( \overline{q^{2}}-\overline{q}^{2}\textrm{ } \),
related to the probability distribution \( \rho  \), and we need to keep track
of it by solving the probability transport equations (\ref{rhot}) in addition
to the modified shallow water system (\ref{hbar}) and (\ref{ubar3}). This
set of equations increases the entropy (at an optimal rate), while preserving
all the conservation laws of the initial inviscid shallow water system. We now
check that the steady solutions reached by these relaxation equations is indeed
the Gibbs state.

\subsection{Relaxation towards the statistical equilibrium}

The entropy production (\ref{Sdot}) can be written 
\begin{equation}
\label{Sdot2}
\dot{S}=-\int {{\mathbf{J}}\over \rho }(\nabla \rho +\beta \rho (\sigma -\overline{q})h\mathbf{u}_{\perp })d^{2}{\mathbf{r}}d\sigma +\beta \int {\mathbf{J}}(\sigma -\overline{q})h\mathbf{u}_{\perp }d^{2}{\mathbf{r}}d\sigma 
\end{equation}
 Using (\ref{Jnorm}) and (\ref{Edot}), the second integral is seen to cancel
out. Substituting for (\ref{Jopt2}) in the first integral, we have 
\begin{equation}
\label{dotS3}
\dot{S}=\int {J^{2}\over D\rho }d^{2}{\mathbf{r}}d\sigma 
\end{equation}
 which is positive provided that \( D\geq 0 \) , and this is clearly a necessary
condition to assure entropy increase in all cases. A stationary solution \( \dot{S}=0 \)
is such that \( {\mathbf{J}}={0} \) yielding, together with (\ref{hu}), 
\begin{equation}
\label{eq1}
\nabla (\ln \rho )+\beta (\sigma -\overline{q})\nabla \psi =0
\end{equation}
 For any reference PV level \( \sigma _{0} \), it writes 
\begin{equation}
\label{eq2}
\nabla (\ln \rho _{0})+\beta (\sigma _{0}-\overline{q})\nabla \psi =0
\end{equation}
 Substracting (\ref{eq1}) and (\ref{eq2}), we obtain \( \nabla \ln ({\rho /\rho _{0}})+\beta (\sigma -\sigma _{0})\nabla \psi =0 \),
which is immediately integrated into 
\begin{equation}
\label{eq4}
\rho ({\mathbf{r}},\sigma )={1\over Z({\mathbf{r}})}g(\sigma )e^{-\beta \sigma \psi }
\end{equation}
 where \( Z^{-1}({\mathbf{r}})\equiv \rho _{0}({\mathbf{r}})e^{\beta \sigma _{0}\psi ({\mathbf{r}})} \)
and \( g(\sigma )\equiv e^{A(\sigma )} \), \( A(\sigma ) \) being a constant
of integration. Therefore, entropy increases until the Gibbs state (\ref{rhopt})
is reached, with \( \beta =\lim _{t\rightarrow \infty }\beta (t) \).

\subsection{Simplified cases:}

In the case of two PV levels \( \sigma _{0} \) and \( \sigma _{1} \), the
transport equation (\ref{rhot}) for the probability \( p_{1} \) is equivalent
to the transport equation for \( \overline{q} \) (since \( \overline{q}=\sigma _{0}+p_{1}(\sigma _{1}-\sigma _{0}) \)),
which is already obtained from the curl of the shallow water equation (\ref{ubar3}).
Therefore the relaxation equations reduce to the modified shallow water system

\begin{equation}
\label{hs}
{\partial h\over \partial t}+\nabla \cdot (h{\mathbf{u}})=0
\end{equation}

\begin{equation}
\label{SHs}
{\partial \mathbf{u}\over \partial t}+\overline{q}h\mathbf{e}_{{z}}\wedge \mathbf{u}=-\nabla ({\mathbf{u}^{2}\over 2}+gh)+D[{\mathbf{e}}_{z}\wedge \nabla \overline{q}-\beta (t)q_{2}h\mathbf{u}]
\end{equation}

\begin{equation}
\label{PVs}
\overline{q}={{(\nabla \wedge \mathbf{u}).\mathbf{e}_{{z}}+2\Omega }\over h}\; \; ,\; \; q_{2}=(\overline{q}-\sigma _{0})(\sigma _{1}-\overline{q})
\end{equation}

\begin{equation}
\label{betas}
\beta (t)=-{\int Dh(\mathbf{e}_{{z}}\wedge \mathbf{u})\nabla \overline{q}d^{2}{\mathbf{r}}\over \int Dq_{2}(\mathbf{e}_{{z}}\wedge \mathbf{u})^{2}h^{2}d^{2}{\mathbf{r}}}
\end{equation}

\begin{equation}
\label{bou}
{\textbf {n}}.\nabla \overline{q}=-\beta (t)q_{2}h{\textbf {n}}{\textbf {u}}_{\perp }\qquad ({\textrm{on }\, \, \textrm{each }\, \, \textrm{boundary}})
\end{equation}

\begin{equation}
\label{bouu}
{\textbf {n}}.{\textbf {u}}=0\qquad ({\textrm{on }\, \, \textrm{each }\, \, \textrm{boundary}})
\end{equation}

where we have omitted the over-bar on \( \mathbf{u} \), and the expression
(\ref{PVs}) of \( q_{2}=\overline{q^{2}}-\overline{q}^{2} \) is easily obtained
for a probability distribution with two values \( \sigma _{0} \) and \( \sigma _{1} \).
The numerical implementation of this system will lead to the two PV levels Gibbs
state.

Stating \( q_{2}=cte \) instead of the expression (\ref{PVs}) yields the Gaussian
Gibbs state with linear relationship between \( \overline{q} \) and \( \psi  \).
Then the coefficient \( q_{2}\beta  \) used in (\ref{SHs}) is directly obtained
from (\ref{betas}). This is sufficient for the purpose of finding the statistical
equilibrium, but more refined relaxation models can be used as discussed by
\cite{ksv98} Kazantsev \textit{et al.} (1998) in the context of QG models.

\subsection{The incompressible limit:}

\label{sec_moment}

The case of ordinary 2D incompressible turbulence is recovered in the limit
\( h\rightarrow 1 \), \( q\rightarrow \omega  \) and \( {\textbf {u}}=-{\textbf {e}}_{z}\wedge \nabla \psi  \).
The relaxation equation (\ref{ubar3}) then becomes 
\begin{equation}
\label{uord}
{\partial \overline{\mathbf{u}}\over \partial t}+(\overline{\mathbf{u}}.\nabla )\overline{\mathbf{u}}=-{1\over \rho }\nabla p+D(\Delta \overline{\mathbf{u}}-\beta (t)\omega _{2}\overline{\mathbf{u}})
\end{equation}
 where we have used the well-known identity of vector analysis \( \Delta \overline{\mathbf{u}}=\nabla (\nabla \overline{\mathbf{u}})-\nabla \wedge (\nabla \wedge \overline{\mathbf{u}}) \)
which reduces for a two-dimensional incompressible flow to \( \Delta \overline{\mathbf{u}}={{\mathbf{e}}}_{z}\wedge \nabla \overline{\omega } \).
Equation (\ref{uord}) is valid even if \( D \) is space dependant unlike with
a usual viscosity term. In previous publications this equation was given only
in its vorticity form 
\begin{equation}
\label{qkt1sqg}
{\partial \overline{\omega }\over \partial t}+\nabla (\overline{\omega }\overline{\mathbf{u}})=\nabla \biggl \lbrack D\biggl (\nabla \overline{\omega }+\beta (t)\omega _{2}\nabla \psi \biggr )\biggr \rbrack 
\end{equation}
 and the equivalence with (\ref{uord}) is not obvious at first sights when
\( D \) is space dependent. At equilibrium, we have from (\ref{uord}) the
identity 
\begin{equation}
\label{id}
\Delta \overline{\mathbf{u}}=\beta \omega _{2}\overline{\mathbf{u}}
\end{equation}
 which can be deduced directly from the Gibbs state. Indeed for a stationary
solution \( \overline{\omega }=F(\psi ) \), the previous identity \( \Delta \overline{\textbf {u}}={\textbf {e}}_{z}\wedge \nabla \overline{\omega } \)
becomes \( \Delta \overline{\mathbf{u}}=-F'(\psi )\overline{\mathbf{u}} \)
equivalent to (\ref{id}) for a Gibbs state thanks to (\ref{q2fprime}).

We now account for a deformation of the fluid layer but assume that the
elevation with respect to the average thickness \( H \) is weak, so that 
\begin{equation}
\label{qg}
h=H(1+\eta )\qquad {\textrm{with}}\qquad \eta \ll 1
\end{equation}
 To first order the flow is incompressible and equation (\ref{h}) reduces to
\( \nabla .{\mathbf{u}}=0 \), or equivalently \( {\mathbf{u}}=-{\mathbf{e}}_{z}\wedge \nabla \psi  \)
(there is a factor \( H \) with the previous definition (\ref{hu})). In the
quasi-geostrophic limit of small Rossby numbers \( \omega \ll \Omega  \), the
momentum equation (\ref{u2}) reduces at zero order to the geostrophic balance
\begin{equation}
\label{uinc}
{\mathbf{u}}={gH\over 2\Omega }{\mathbf{e}}_{z}\wedge \nabla \eta \qquad {\textrm{or}}\qquad \psi =-{gH^{2}\over 2\Omega }\eta 
\end{equation}
 The expression for the potential vorticity then reduces to 
\begin{equation}
\label{qQG}
\zeta \equiv Hq-2\Omega \simeq \omega +{\psi \over L_{R}^{2}}
\end{equation}
 with the Rossby radius of deformation 
\begin{equation}
\label{Rossby}
L_{R}={\sqrt{gH}\over 2\Omega }
\end{equation}
 The term \( {1\over L_{R}}\psi  \) in (\ref{qQG}) creates a shielding of
the interaction between vortices (similar to the Debye shielding in plasma physics)
on a length scale \( \sim L_{R} \). In the limit \( 1/L_{R}\rightarrow 0 \),
we recover the 2D incompressible equations. For finite \( L_{R} \) we can extend
the statistical theory of the 2D Euler equations to the QG case by simply replacing
the vorticity \( \overline{\omega } \) by the potential vorticity \( \overline{\zeta } \)
\cite{sndr91,mr94,ksv98,bs00}.

\section{The case of circular domains or channel:}

\label{sec_geometries}

\subsection{Statistical equilibrium}

In a disk, the angular momentum

\begin{equation}
\label{L}
L=\int h(\mathbf{r}\wedge \mathbf{u})_{z}d^{2}\mathbf{r}
\end{equation}
 is conserved. This additional constraint can be accounted for by adding a term
\( \beta \lambda \delta L \) in the first order variation (\ref{varia}). We
can write \( \delta L=\int \delta h(\mathbf{r}\wedge \mathbf{u})_{z}d^{2}\mathbf{r}+\int h(\mathbf{e}_{{z}}\wedge \mathbf{r}).\delta \mathbf{u}d^{2}\mathbf{r} \),
and the second term can be expressed in terms of \( \delta \overline{\omega } \)
and \( \delta (\nabla .\mathbf{u}) \) by a Helmholtz decomposition of \( h(\mathbf{e}_{{z}}\wedge \mathbf{r}) \)
analogous to (\ref{hu}), followed by an integration by part. This is analogous
to the formulae (\ref{dE2})(\ref{dE3}) used for expressing the energy variation.
We can combine the energy and momentum variations by defining 
\begin{equation}
\label{comb}
h[{\mathbf{u}}-\lambda (\mathbf{e}_{{z}}\wedge \mathbf{r})]=-{\mathbf{e}}_{z}\wedge \nabla \psi '+\nabla \phi '
\end{equation}
 instead of (\ref{hu}). Adding the new terms in (\ref{var1})(\ref{var3})
yields the Gibbs state (\ref{rhopt})(\ref{Bpsi}) for the velocity \( \mathbf{u}'={\mathbf{u}}-\lambda (\mathbf{e}_{\mathbf{z}}\wedge \mathbf{r}) \)
seen in the reference frame rotating at angular velocity \( \lambda  \). Accordingly,
the expression of the Bernouilli function must be modified by a term of centrifugal
force: we must use \( B'(\psi ')=gh+{{\mathbf{u}'}^{2}\over 2}-\lambda ^{2}r^{2} \)
instead of \( B(\psi ) \). We find therefore that the Gibbs state (its locally
averaged velocity field) is a solution of the shallow water equation which is
steady in a reference frame rotating at a modified angular velocity \( \Omega +\lambda  \).
This velocity is indirectly determined by the constraint on angular momentum.
Note that the result can be readily extended to the shallow water system on
the sphere.

In the case of an annulus, the circulation \( C_{-}=-\int u_{\theta }dl \)
around the inner wall is an additional conserved quantity (the circulation \( C_{+} \)
around the outer wall is also conserved, but it is related to other conserved
quantities, as \( C_{+}=\Gamma -C_{-} \), and the conservation of \( \Gamma  \)
is already included in the PV conservation). Furthermore we need in general
to set \( \psi =\psi _{-}\neq 0 \) at the inner wall (while we can still set
\( \psi =0 \) at the outer wall). As a consequence a boundary term -\( \psi _{-}\delta C_{-} \)
now appears in the expression (\ref{dE4}) for the energy variation. However
we can directly set \( \delta C_{-}=0 \), canceling this boundary term, without
influence on the independent variables \( h\rho  \) (determining the locally
averaged vorticity \( \overline{\omega }=\int \sigma h\rho d\sigma  \)), \( \nabla .\mathbf{u} \)
and \( h \). Therefore the only modification with respect to the disk is an
additional unknown \( \psi _{-} \) in the definition (\ref{psi}) of \( \psi  \),
determined by the corresponding additional constraint \( C_{-} \) .

The case of a straight channel can be viewed as the limit of an annulus with
small width, but it can be convenient to treat it in itself. Let us consider
a straight channel between two walls at coordinates \( y=\pm L_{y}/2 \) with
periodic boundary conditions along the x-direction (with domain length \( L_{x} \)
). In the absence of Coriolis force (\( \Omega =0) \) , the x-wise momentum
\begin{equation}
\label{P}
P=\int hu_{x}d^{2}\mathbf{r}
\end{equation}
 is conserved (instead of the angular momentum), as well as the circulation
\( C_{+}=-\int u_{x}dx \) along the upper wall (\( y=L_{y}/2 \) ). The boundary
condition (\ref{psi}) defining \( \psi  \) is replaced by \( \psi =P/L_{x} \)
at the upper wall \( y=L_{y}/2 \) and \( \psi =0 \) (for instance) at the
lower wall \( y=-L_{y}/2 \). Unlike with angular momentum, we cannot express
the variation \( \delta \mathbf{P} \) in terms of the variations in the independent
variables \( \overline{\omega } \) , \( \nabla .\mathbf{u} \) , \( h \).
However we have now an additional freedom in the variational problem, as we
can add a uniform x-wise velocity \( -U\mathbf{e}_{{x}} \) (use a reference
frame with velocity \( U\mathbf{e}_{{x}} \)) without influence on the independent
variables \( \overline{\omega } \) , \( \nabla .\mathbf{u} \) , \( h \).
For any choice of \( U \), we can solve the variational problem with the velocity
\( \mathbf{u}'=\mathbf{u}-U\mathbf{e}_{{x}} \) and corresponding energy \( E'=E+MU^{2}/2-PU \),
upper wall circulation \( C_{+}^{_{\prime }}=C_{+}-UL_{x} \). This yields a
Gibbs state (\ref{rhopt})(\ref{qpsi})(\ref{Bpsi}), representing a steady
solution of the shallow water equation in a reference frame moving with velocity
\( U\mathbf{e}_{\mathbf{x}} \). Among these states, the ones with the right
value of the momentum \( P=\int hu_{x}d^{2}\mathbf{r} \) will be the actual
solutions. Families of Gibbs states with the same structure translated in the
x-direction are obtained, as discussed by \cite{ssr91} Sommeria \textit{et al.}
(1991) in the incompresible case.

Finally in the case of an infinite domain, the two components of momentum, as
well as the angular momentum are conserved. This yields to purely translating
or purely rotating Gibbs states, as discussed by \cite{csom} Chavanis and Sommeria (1997)
in the incompressible case.

\subsection{Relaxation equations}

Taking the derivative of (\ref{P})(\ref{L}) with respect to time and using
equations (\ref{hbar})(\ref{u2bar}), we obtain the constraints 
\begin{equation}
\label{Pdot}
\dot{P}=\int h{{J}}_{\omega y}d^{2}{\textbf {r}}={{0}}
\end{equation}
\begin{equation}
\label{Ldot}
\dot{L}=-\int h{\textbf {J}}_{\omega }.{\textbf {r}}d^{2}{\textbf {r}}={ {0}}
\end{equation}
 These constraints can be included in the variational principle (\ref{vp})
by introducing appropriate Lagrange multipliers denoted \( \beta (t){{U}}(t) \)
and \( \beta (t){\lambda }(t) \). Then, the results of section \ref{sec_relaxation}
are generalized simply be replacing the velocity \( \overline{\textbf {u}} \)
by the relative velocity \( \overline{\textbf {u}}'=\overline{\textbf {u}}-{{U}}(t){\textbf {e}}_{x}-{\lambda }(t){\textbf {e}}_{z}\wedge {\textbf {r}} \)
where the time evolution of \( {{U}}(t) \) and \( {\lambda }(t) \) is obtained
by substituting the optimal current (\ref{Jomega}), with the above transformation,
in the constraints (\ref{Pdot}) and (\ref{Ldot}). In the case of a channel
we have the additional conserved quantity \( C_{+}=-\int u_{x}dx \) along the
upper wall. Using (\ref{u2bar}), we readily find that \( \dot{C}_{+}=\int J_{\omega y}dx=0 \)
as the current \( \mathbf{J}_{\omega } \)is parallel to the wall, so the circulation
along each wall is indeed conserved by the relaxation equations.

\end{document}